\documentclass[preprint,12pt]{elsarticle}

\usepackage{amssymb}
\usepackage{amsmath}

\usepackage{color}

\journal{Energy and Buildings}

\begin{document}

\begin{frontmatter}



\title{Optimal chiller loading including transients}


\author[inst1]{Manuel R. Arahal}
\author[inst1]{Manuel G. Satu\'e}
\author[inst1]{Manuel G. Ortega}

\affiliation[inst1]{organization={Systems Engineering and Automation Department},
            addressline={ETSI, U. Sevilla}, 
            city={Seville},
            postcode={41092}, 
            state={},
            country={Spain}}


\begin{abstract}
Scheduling and loading of chillers in a multi-chiller plant is considered. A new framework is introduced considering an extended set of independent variables for the optimization problem of energy consumption. In this way the number of decision variables is increased, providing extra degrees of freedom to optimize cooling plant operation. The dynamic effects due to transients arising from switching on and off of units are usually not considered in the literature dealing with Optimal Chiller Loading/Sequencing which is restricted to the static case. In this paper, these effects are treated in a way that results in  a manageable optimization problem.  A Simultaneous Perturbation Stochastic Approximation solution is deployed for the problem and the proposed method is compared with a similar but static approach showing the benefits in terms of reduced energy consumption. 
\end{abstract}

\begin{graphicalabstract}
\begin{tabular}{|p{12cm}|} \hline
1. Present Optimal Chiller Loading/Sequencing as a dynamic optimization problem \\
2. Develop mathematical models to include transients \\
3. Define an extended set of decision variables for the problem \\
4. Illustrate the SPSA approach to the extended dynamic Optimal Chiller Loading/Sequencing \\
5. Compare results with static case in different scenarios \\
\hline
\end{tabular}
\end{graphicalabstract}

\begin{highlights}
\item Transients are considered in the Optimal Chiller Loading/Sequencing problem in a way that the resulting optimization problem is tractable
\item Decision variables are extended with respect to previous OCL/OCS approaches
\item SPSA is deployed and hyper-parameters tuning is discussed 
\end{highlights}

\begin{keyword}
Chiller \sep Economic dispatch \sep  Optimization \sep Sequencing \sep Transients
\end{keyword}

\end{frontmatter}


\section{Introduction}
\label{sec:sample1}
\textcolor{black}{The design of cooling systems often uses a multiple-chiller option since it provides flexibility, stand-by capacity and better maintenance. Also, during operation, the multi-chiller option provides reduced starting-up electrical peak demand and better performance under partial-load conditions \cite{dossat1961principles}. During design, peak thermal load is considered;} then, during operation, the actual cooling load is below the design limit and it is not constant (i.e. time-varying). In such circumstances, it is convenient to resort to some means to partition the load among the units in the multi-chiller plant. Optimal Chiller Loading (OCL) aims at producing such partition  taking into account an economic objective (i.e. reducing costs) \cite{chang2004novel}. \textcolor{black}{Many optimization schemes have been proposed for the OCL problem, in particular for the  interesting  case where the chiller units at the plant have different characteristics \cite{salari2015new}, producing a non-convex optimization problem. This requires the use of computing time intensive algorithm compared with the convex situation. The lack of convexity arises mainly from the characteristic curves of the chillers relating consumption to part load ratio \cite{chang2005genetic}.}

\textcolor{black}{The basic OCL problem \cite{chang2004novel,salari2015new} can be solved at any given time regardless of past or future information.} This is done considering steady-state relationships and neglecting dynamic effects such as energy storage and transients. A very different scenario arises when the optimization takes into account dynamic effects. The simplest of these scenarios might be the inclusion of a penalization on the number of switch changes in the state of the chillers. Such penalization can be used to prevent frequent changes in the on/off state of the chillers that are not designed to operate in such way. In this new scenario, the state at one given time influences the OCL problem for subsequent time instants. Then, the problem of Optimal Chiller Sequencing (OCS) appears where a temporal horizon (e.g. a day or a week) must be taken into account as a whole.

Some papers have considered the economic dispatch for multi-chiller plants within the static framework introduced above. For instance, in \cite{lo2016economic}  a ripple bee swarm optimization algorithm is used to produce weekly scheduling for six chillers having a non-convex kW-PLR function such as those of the Hsinchu benchmark \cite{chang2005genetic}. Cooling demand forecasts with uncertainties have been considered in \cite{saeedi2019robust}, where different optimization methods are compared. In \cite{shao2019chiller}, the different electricity prices during the day are also considered to further reduce costs. The hour-ahead optimization by means of imperialistic competitive algorithm is proposed in \cite{jabari2020energy}.

The optimization needed for the OCS is an optimal control problem as it uses a cost functional that is a function of state and control variables. Other papers have also considered this kind of problem. For instance, in  \cite{acerbi2020exact} minimum up-time and down-time requirements on chillers’ operation are considered.  Other papers have also faced the optimal control problem due to the use of dynamic components such as ice-storage \cite{chen2005optimization,deng2014model}. A related approach that can deal with the scheduling problem is that of predictive control. There, a moving window provides a future framework where control actions are selected to optimize a functional, such as in \cite{candanedo2013model}. Please note that the receding horizon strategy does not provide (in general) a global optimal solution \cite{camacho2013model}. Whatever the case, for the OCS to be tackled a model to link control variables to consumption is needed. In most cases  data-sheets from manufacturers are used, in other cases this data is enhanced with simulations using tools such as TRNSYS \cite{wang2018practical}. Data-driven approaches have also been proposed in this context  \cite{wei2014modeling,sala2020predictive,ho2021variable}.

\textcolor{black}{In the OCL/OCS literature, the power load of each chiller in the plant is considered as a way to optimize consumption. This variable, termed Part Load Ratio (PLR), is used as an independent or decision variable. However, the PLR is usually not directly controlled during operation, instead it is a variable resulting from the combination of mass flows and temperature set-points for the chillers that, in turn, are manipulable. Also, it has been reported that  maximum capacity of chillers can be increased altering the input and output temperatures of water in the chiller \cite{huang2016amelioration}. These observations have made some researchers to consider other variables as potential opportunities to reduce the power consumption of a cooling plant. In this paper, the mass flow rate and the output temperature reference for each chiller are considered,  similarly to the work in \cite{thangavelu2017energy, chiam2019hierarchical} but considering transients. Also in some papers, the number of units that are to be used at any given time (active chillers) is separately optimized from their loading (see \cite{huang2016amelioration} and the numerous references therein). Here, the state (on/off) is concurrently used with the rest of the optimization variables in a way similar to the holistic optimization proposed in \cite{karami2018particle}.}

\textcolor{black}{Another aspect to be considered is the power consumption of the secondary pump. In the above cited literature this is not taken into consideration as part of the optimization problem. In fact,  mass flows are usually prefixed so that to cover the cooling demand and then the PLR is optimized \cite{chang2005genetic,saeedi2019robust,shao2019chiller,jabari2020energy}. In the present work,   mass flow rates are also part of the independent variable, hence the consumption of the secondary pump will be included separately as done in \cite{karami2018particle}. Similarly, in all of the above cited papers, thermal losses are considered to be part of the whole cooling load. However, the chilled water temperature is considered here as a variable. It is well known that thermal losses depend (among other factors) on chilled water temperature, hence thermal losses will be considered as another variable entering the optimization problem.}

In addition to the above innovative contribution, in this paper, the problem of costs associated with transients is considered. \textcolor{black}{Transients are a source of losses since during operation chiller have some thermal energy stored (thermal and mechanical) that is lost after a shut-down and must be provided again during a later start-up. In \cite{lu2015optimal}, a fixed penalization for the switching operations is used. Instead, here the dynamic effects associated with the transients are modelled, so there is no need to introduce artificial penalties. The model is developed in a way that the resulting optimization problem is manageable.} A Simultaneous Perturbation Stochastic Approximation (SPSA) \cite{spall1992multivariate} solution to the OCS problem is then proposed as a means for dealing with the high number of decision variables. 

The next section presents with more detail the dynamic OCS problem that constitutes the object of this paper. In Section 3 the proposed method is presented, modeling the dynamic effects in a way amenable to the OCS problem. Also the SPSA algorithm is deployed for the dynamic OCS. The results on a number of scenarios are presented in Section 4 and compared with those attained using the static method.

\section{Problem statement}
The economic dispatch or OCS problem in the presence of dynamic effects such as transients is discussed in the following. It will be shown that a mathematical model of the cooling plant can be derived, thus allowing to compute the consumption for a given sequence and loading of the chiller units. This, in turn, allows for the optimization of consumption by carefully selecting the on/off sequence, the mass flow rates and the temperature set-points. This optimization will also ensure that the thermal load is satisfied, including losses. 

\textcolor{black}{The problem is to decide for each time period in a given horizon (e.g. 24 hours in a day) the working point for each chiller in a cooling plant. Thus, discrete-time variables are considered being the sampling time $T_s$ usually equal to one hour.} The working point is defined by the set point for the output temperature $T^o$ (K) and by the mass flow rate $\dot{m}$ (Kg/s). The objective is to provide enough cooling power for the demand and losses $\dot{Q}_L$ with the minimum possible electrical consumption. A forecast of the variables that are not under control (e.g. building occupancy, external temperature) is needed \cite{en14123479}. Thus, the main ingredients of the OCS problem are the plant model, the predictions and a means to solve the optimization problem.

The electrical consumption of a plant with $N$ chillers is defined as

\begin{equation}
    \label{eq_consumototalP}
    P = \sum_{i=1}^N P_i + P_s + P_e
\end{equation}
\noindent where $P_i$ is the electrical power drawn by chiller $i$ (including power for ancillary devices such as control electronics, fans, etc), $P_s$ is the electrical power drawn by the secondary pump and $P_e$ is the extra electrical power needed for transients and other effects that will be discussed later.   

With the above definitions, the optimization problem posed by the OCS is defined in mathematical form as
\begin{equation}
    \label{eq_optimz_probl}
\begin{aligned}
\min_{ {\bf x}} \quad & P( {\bf x}) \\
\textrm{s.t.} \quad &  \dot{Q}( {\bf x})  = \dot{Q}_L \\
                    & x_{i} \le {\bf x}  \le x_{s} \\
                    &  g_{i} \le G( {\bf x})  \le g_{s}, 
\end{aligned}
\end{equation}

\noindent \textcolor{black}{where the independent variable ${\bf x}$ is a vector containing the discrete-time trajectories from $t=1$ to $t=N_T$ of the variables that define the working point for each chiller $i$ with $i=1, ..., N$. In mathematical form,}

\begin{equation}
    \label{eq_comp_x}
    {\bf x} = \left( \dot{m}_s |_1^{N_T},  ~ \dot{m}_1 |_1^{N_T},  ~ T_1^o |_1^{N_T}  , ~ ..., \dot{m}_N |_1^{N_T},  ~ T_N^o |_1^{N_T}    \right)^\intercal
\end{equation}

\noindent where $\dot{m}_s$ is the mass flow rate for the secondary pump that is also an independent variable due to the decoupling provided by the by-pass pipe. Please note that the sign $|_1^T$ denotes a set of points within a trajectory. For a generic variable $v$ the following definition can be given
\begin{equation}
    \label{eq_traject}
    v |_1^{N_T} \doteq \left( v(1),  ~ v(2),  ~ ...,  ~ v(N_T) \right)^\intercal.
\end{equation}

In the definition of the optimization problem of (\ref{eq_optimz_probl}), the following restrictions are used:

\begin{itemize}
    \item $\dot{Q}( {\bf x})  = \dot{Q}_L$ this ensures that the cooling power delivered by the chiller units covers the demand (and losses) for each time period. The time index has been omitted for clarity.
    \item $x_{i} \le {\bf x}  \le x_{s}$ this ensures that the independent variables stay within the particular limits imposed by the installation. 
    \item $g_{i} \le G( {\bf x})  \le g_{s}$ this is a set of constraints expressed not in terms of $x$ but in terms of functions of $x$. $G$ is a vector function with several elements that allow to include  other requirements for instance, in the power consumption for each chiller (limited by the installation) or in the temperature drop of the water across each chiller. 
\end{itemize}

The problem of (\ref{eq_optimz_probl}) is a multivariate general optimization problem in which the independent variable has dimension $\dim{  {\bf x} } = T \cdot ( 2N+1)$ with $N$ being the number of chiller units and $N_T$ the number of time-steps. In many papers $N_T$ ranges in $[8, 24]$ (hours) and the number of chillers varies between 3 and 9. Thus, even for a small installation the dimension of ${\bf x}$ is no less than 50. The proposed optimization algorithm can handle this type of optimization problem needing just forward models to compute the objective function $P$ and the constraints.

\subsection{Plant setup and models}
\begin{figure}
    \centering
    \includegraphics[width=9 cm]{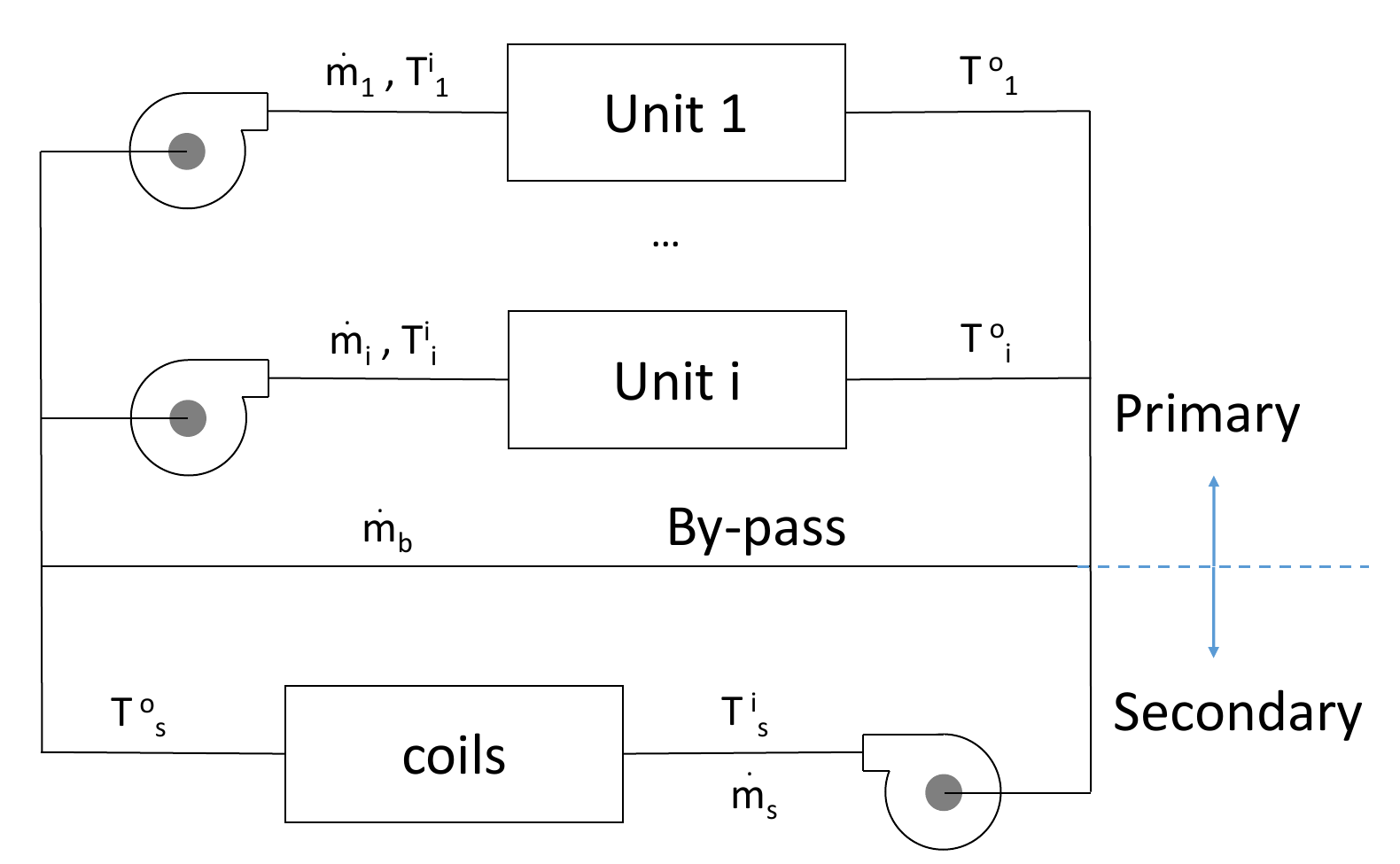}
    \caption{Diagram of a multi-chiller cooling plant connected to a load.}
    \label{fig_diagr}
\end{figure}

The multi-chiller plant considered in this work consists of several chillers connected in parallel with the load and with a by-pass pipe as shown in Figure \ref{fig_diagr}. \textcolor{black}{Different model types (white-box, black-box, grey-box models) have been used for control as surveyed in \cite{li2014review}. White-box models are based on physics and underlying principles, thus providing accurate predictions \cite{RODRIGUEZ2020114415}. For some processes white-box modelling is difficult and one must resort to grey or black box approaches. Grey-box models combine an inner structure of the system dynamics  with the fitting of some free parameters to data in a way similar to that of systems identification \cite{arahal2008serial}}. In  this paper,  first principles structures with parameters adjusted from manufacturer data-sheets are used, as is the case of chillers and pumps, producing grey-box models. The thermal load due to building occupancy (or installation usage) and ambient temperature are forecasts made using black-box approaches as done in \cite{en14123479}.

\textcolor{black}{The modeling begin providing the following expression for the cooling by the $i$-th chiller unit in the plant}

\begin{equation}
    \label{eq_qpi}
    \dot{Q}_i = \dot{m}_i c_p (T_i^i-T_i^o),
\end{equation}

\noindent where $\dot{Q}_i$ is the cooling power (W), $\dot{m}_i$ is the mass flow rate (kg/s), $c_p$ is the specific heat of water, $T_i^i$ is the temperature of the water entering the unit (K) and $T_i^o$ is the temperature of the water leaving the unit (K). The part load ratio (PLR) for the unit is defined as 

\begin{equation}
    \label{eq_plr}
    PLR_i = \frac{\dot{Q}_i}{\dot{Q}_i^n},
\end{equation}

\noindent where $\dot{Q}_i^n$ is the nominal cooling power of the unit. The coefficient of performance (COP) for the chiller is obtained from data provided by the manufacturer. Th and it is a function f$(.)$ of several variables as follows:

\begin{equation}
    \label{eq_cop}
    COP_i = \mathrm{f} \left( T_i^{ev}, T_i^{co}, PLR_i \right),
\end{equation}

\noindent where $T_i^{ev}$ is the temperature at the evaporator (K), $T_i^{co}$ is the temperature at the condenser (K) and $PLR_i$ is the part load ratio already defined. In this coefficient all power consumed at the chiller is included as losses, including power for the control electronics. In this way it is possible to compute 

\begin{equation}
    \label{eq_kwi}
    P_i = \frac{\dot{Q}_i}{COP_i}
\end{equation}
where $P_i$ (W) is the electrical power to be provided to the unit in order to produce the desired cooling. The temperature of the water entering the secondary can be obtained by the following expression
\begin{equation}
    \label{eq_Tsi} 
    T_s^i = \frac{ \sum_{i=1}^{N} \dot{m}_i T_i^o }{ \sum_{i=1}^{N} \dot{m}_i },
\end{equation}
\noindent where, $\dot{m}_p$ is the mass flow rate for the primary defined as 

\begin{equation}
    \label{eq_mp} 
    \dot{m}_p = \sum_{i=1}^{N} \dot{m}_i.
\end{equation}

The thermal load due to occupancy $\dot{Q}_o$ (W) and due to thermal losses $\dot{Q}_l$ (W) produce an increment in the temperature of water that returns from the secondary at temperature $T_s^o$ (K) such that

\begin{equation}
    \label{eq_Tso} 
   \dot{Q}_o + \dot{Q}_l = \dot{m}_s c_p ( T_s^o - T_s^i).
\end{equation}

In the above equation $\dot{m}_s$ is the mass flow rate of water at the secondary (kg/s) and can be computed as

\begin{equation}
    \label{eq_mps} 
   \dot{m}_s = \dot{m}_p - \dot{m}_b
\end{equation}

\noindent where $\dot{m}_b$ is the mass flow rate in the by-pass. Finally, temperature $T_i^i$ can be computed from 

\begin{equation}
    \label{eq_Tii} 
    T_i^i = \frac{ \dot{m}_b T_s^i + \dot{m}_s  T_s^o }{  \dot{m}_p },
\end{equation}

The values of COP for each chiller come from the manufacturer in the form of several 2D tables and must be interpolated in order to produce function f$(.)$ in equation (\ref{eq_cop}).


\section{Proposed method}
The scenario considered so far is most often used case in the literature dealing with OCL and OCS. It does not include dynamic effects and some aspects are also simplified as will be shown next. In this paper the scenario considered deviates from the previous one in two respects:  1) transients due to on/off operation of chillers will be treated as extra losses, and 2) secondary pump operation and thermal losses will be considered not as part of the load demand but separately as they depend on the operating conditions. In the following these two aspects are discussed.

\subsection{Integration of dynamic effects}
\begin{figure}
    \centering
    \includegraphics[width=12cm]{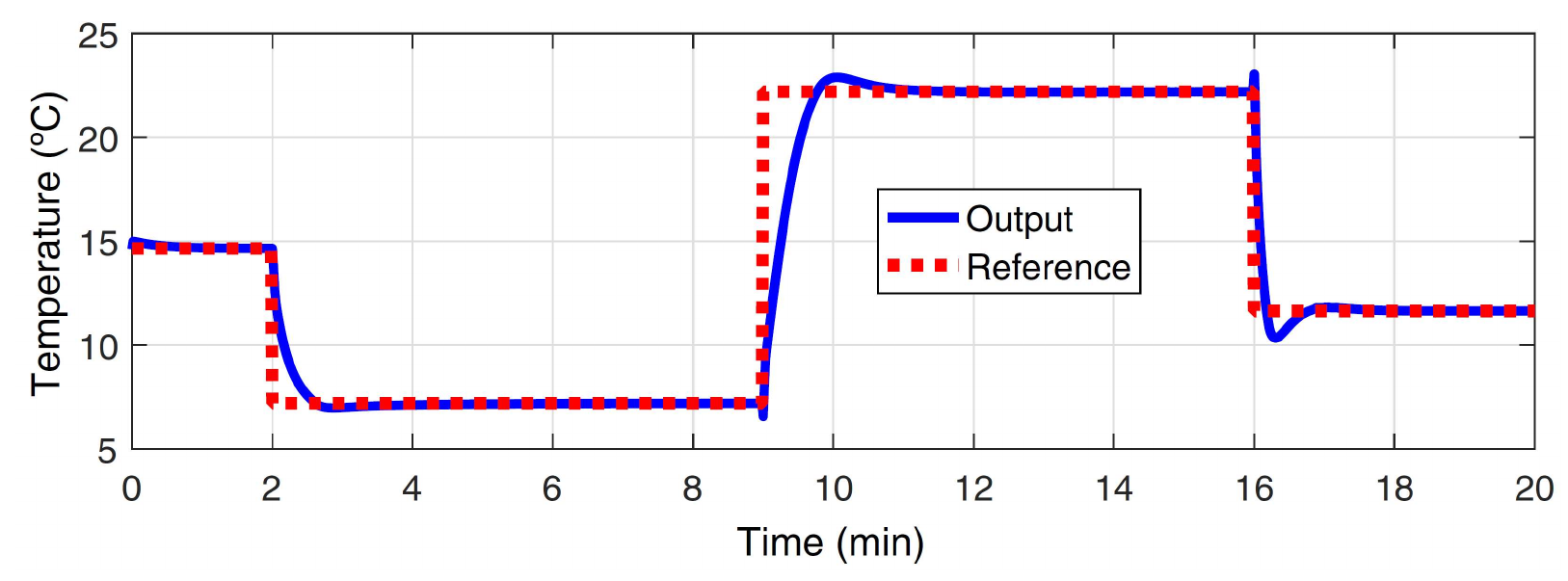}
    \caption{Typical transient for the temperature of the refrigerant gas at the evaporator.}
    \label{fig_graf_transient}
\end{figure}

\textcolor{black}{Transients in chillers have been studied  considering first principles equations of mass and energy balance. With certain conditions for the flow of fluids (e.g. Reynolds number) it is possible to develop transient simulations that can be checked against real data.  From such modelling it is apparent that the dynamics can be approximated by a set of first order transfer functions connected in series. See for instance the reported results in \cite{lei2005dynamic,yao2015state} to cite a few.  Refrigerant charge migration, thermal capacitance of the components, and energy storage in the refrigerant contained in the chiller and adjacent pipes all can have different time constants depending on the installation}. The resulting effect is a dynamic response that can be modelled using black box models such as transfer functions. In Figure \ref{fig_graf_transient} a typical response is shown for the temperature of the refrigerant gas at the evaporator taken from \cite{bejarano2018benchmark}. A similar approach is  used in \cite{romero2011simplified}, where discrete-time transfer functions 
are identified from data and in \cite{yao2015state} where a continuous-time state-space formulation is used.

It can be seen that a first or second order approximation can be obtained. The values reported in the literature for the time constant are well below the usual time-step of one hour considered in the OCL/OCS problems. 

\subsection{SPSA algorithm and implementation}
The Simultaneous Perturbation Stochastic Approximation (SPSA) method can solve optimization problems such as the one posed in (\ref{eq_optimz_probl}). It is an iterative method that relies on an stochastic estimation of the gradient of the objective function. An interesting feature of the SPSA is that the gradient approximation requires only two samples of the objective function for any value of $\dim( {\bf x})$. This makes it an attractive method for large-scale problems.

At any given iteration $j$, the tentative solution ${\bf x}_j$ is computed as

\begin{equation}
    \label{eq_iterative_x} 
    {\bf x}_j = {\bf x}_{j-1} - a_j \hat{J}( {\bf x}_{j-1} ),
\end{equation}

\noindent where $a_j$ is the step size for the current iteration and $\hat{J}$ is an estimation of the Jacobian vector of the objective function defined as

\begin{equation}
    \label{eq_jacobian} 
    J = {\frac {\partial P}{\partial {\bf x}}}.
\end{equation}

The approximation of the gradient is computed using a random perturbation vector $\Delta$ of the same dimension as ${\bf x}$. Then for each component $k$ 

\begin{equation}
    \label{eq_grad_aprox}
    \hat{J}_k( {\bf x}_{j-1} ) = \frac{ P( {\bf x}_{j-1} + c_j \Delta_k  ) - P( x_{j-1} - c_j \Delta_k  )} {2 c_j \Delta_k}.
\end{equation}

In the above definitions, the sequences $\{ a_j \}$ and $\{ c_j \}$ are monotonically decreasing with $j$. The actual values affect convergence and must be chosen as will be discussed later. The iterations  provide a decreasing sequence of objective function values $\{ P( {\bf x}_j) \}$ whose limit is a minimum.

In this paper some work has been done to tune the SPSA method to the particular problem at hand. In particular as means of obtaining the sequences $\{ a_j \}$ and $\{ c_j \}$  has been derived. The final implementation has the following steps:

\begin{enumerate}
  \item Set $j=0$ and apply initial value ${\bf x}_0$.
  \item Compute the gains as $a_j = a / (A+j)^\alpha$ and $c_j = c/j^\gamma$.
  \item Compute the random perturbation vector $\Delta_j$ by sampling a Bernoulli $\pm 1$ independent distribution with a probability of 0.5 for each possible outcome. 
  \item Evaluate the objective function and approximate the gradient using (\ref{eq_grad_aprox})
  \item Update the estimate by means of (\ref{eq_iterative_x})
  \item Test the feasibility of the points ${\bf x}_{j}$, if constraints are violated then go back to 3.
  \item If the termination condition is met then finish, else go back to step 2.
\end{enumerate}

The initial feasible solution ${\bf x}_0$ and the sequences  $\{ a_j \}$ and $\{ c_j \}$  are discussed in the following.

\begin{enumerate}
  \item The initial guess $x_0$  \textcolor{black}{can be set by considering a chiller load proportional to their capacity.}
  \item Hyper-parameters $A$, $a$, $\alpha$, $c$ and $\gamma$ define the
  sequences that govern the step size at each iteration and the magnitude
  of the perturbation, respectively. The values of the
hyper-parameters used are $a=0.16$, $A=200$, $\alpha=0.6$,
$c=0.35$, $\gamma=0.1$, $k_{min} = 20$, $k_{max}= 10^3$ and
$\beta=0.01$.

  \item The termination condition is usually posed in terms of the
  variation of the current estimate ${\bf x}_{j}$ with respect to
  the previous one and can be computed as
  $\| {\bf x}_j - {\bf x}_{j-1} \|  < \epsilon$. The value of vector
  $\epsilon$  is obtained as a fraction $\beta$ of $\| {\bf x}_1 -
{\bf x}_j \|$ for $j>j_{min}$, being $j_{min}$ a minimum number of
iterations that the algorithm is enforced to go through. 
\end{enumerate}

\textcolor{black}{The SPSA method provides a minimum of the objective function. As happens with many optimization methods for nonlinear problems, there is no guarantee of a global optimization. Although the algorithm possesses some capabilities to escape local minima there is no guarantee
that it will always evade them. In other words, the minimum might be a local one. Unfortunately, there is not yet an algorithm to produce the global minimum for each case. As a result, the performance evaluation cannot be produced. Instead a distribution of the local minima found by repeatedly solving the problem with different initial conditions and different parameters can be produced to assess the convergence of the algorithm as shown in the next Section.}

\section{Experimental results}
The proposed method will be compared against two common procedures: sharing of load proportional to the chiller's capacity and OCL. In both cases dynamic effects are not taken into account. The methods considered will be put to test in different scenarios characterized by different trajectories of cooling load $\dot{Q}_L(t)$ and ambient temperature $T_{amb}(t)$. In all cases the methods are required to fulfill the demand and associated thermal losses maintaining the operational variables within limits.

A multi-chiller plant with thee units ($N=3$) is considered to illustrate the proposed methodology and to compare it with previous approaches. Three different commercial TRANE air-cooled chillers are used for the reported experiments: RTAC models 400STD, 300STD and 250STD. Please note that the methodology is independent of this choice and can be extended to any number of units with different characteristics. The main characteristics of the chillers are presented in Table \ref{tab_trane}.

\begin{table}[htp]
    \centering
    \begin{tabular}{llllll} 
        Model   & Power  & $COP^*$  & $\dot{m}_{min}$ & $\dot{m}_{max}$ & IPLV \\
                & (RT) &         & ($kg/s$)        &  ($kg/s$) & \\ \hline
        400 STD & 400   &  2.8     &  29             & 105      & 13.7 \\
        300 STD & 300   &  2.8     &  20             &  68      & 13.7 \\
        250 STD & 250   &  2.8     &  16             &  59      & 12.8 \\
    \end{tabular}
    \caption{Main characteristics of the chillers. $COP^*$ is given for full load at nominal conditions. IPLV values are rated in accordance with ARI Standard 550/590-98.}
    \label{tab_trane}
\end{table}

\begin{figure}
    \centering
    \includegraphics[width=9 cm]{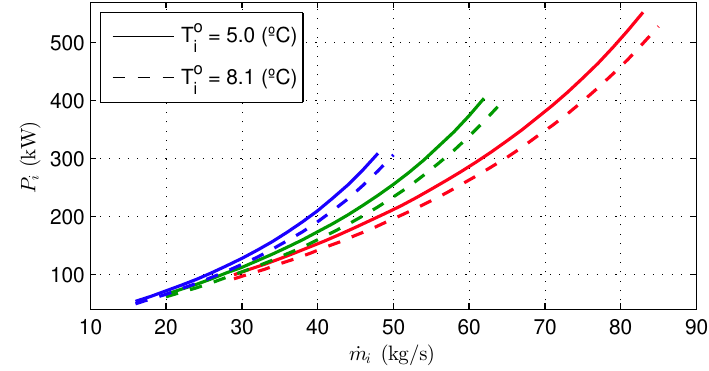}
    \includegraphics[width=9 cm]{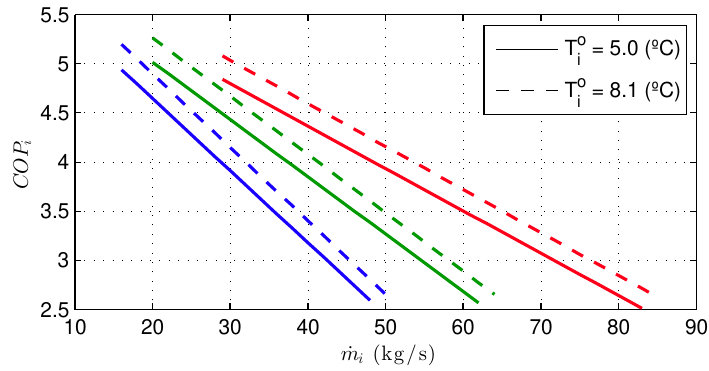}
    \caption{Electrical power and COP for different mass flow rates and two output temperatures. Color indicates chiller's model: red for 400STD, green for 300STD, and blue for 250STD. Line style indicates output temperature $T^o_i$ as shown in the inset.}
    \label{fig_caract_trane}
\end{figure}

As an example, Fig. \ref{fig_caract_trane} shows the electrical power $P_i$ and COP as a function of mass flow rate for each model considered in the plant.

\textcolor{black}{The SPSA can easily be incorporated by different programming languages. In this case MATLAB has been used for ease of implementation. In \cite{spall1992multivariate} some programming tips are provided to help coding the SPSA method in other programming languages.}

\subsection{Solutions for different scenarios}
\textcolor{black}{The scenarios considered are for a large university hospital with several modules served by different cooling plants. In this module, the occupational load is due mainly to hospital staff and patients arriving during the day for diagnosis.} Three different scenarios (TL1, to TL3) will be considered. Each one is characterized by a particular ambient temperature trajectory $T_{amb}(t)$ and load profile $\dot{Q}_L(t)$. The  load profiles correspond to 2 working days with different occupancy profiles plus another day with reduced occupancy concentrated in a few hours corresponding to a half-holiday. In all cases 12 hours will be studied (from 7am to 19pm) as the occupancy of the installation outside those hours is almost null.

\begin{figure}
    \centering
    \includegraphics[width=6.6cm]{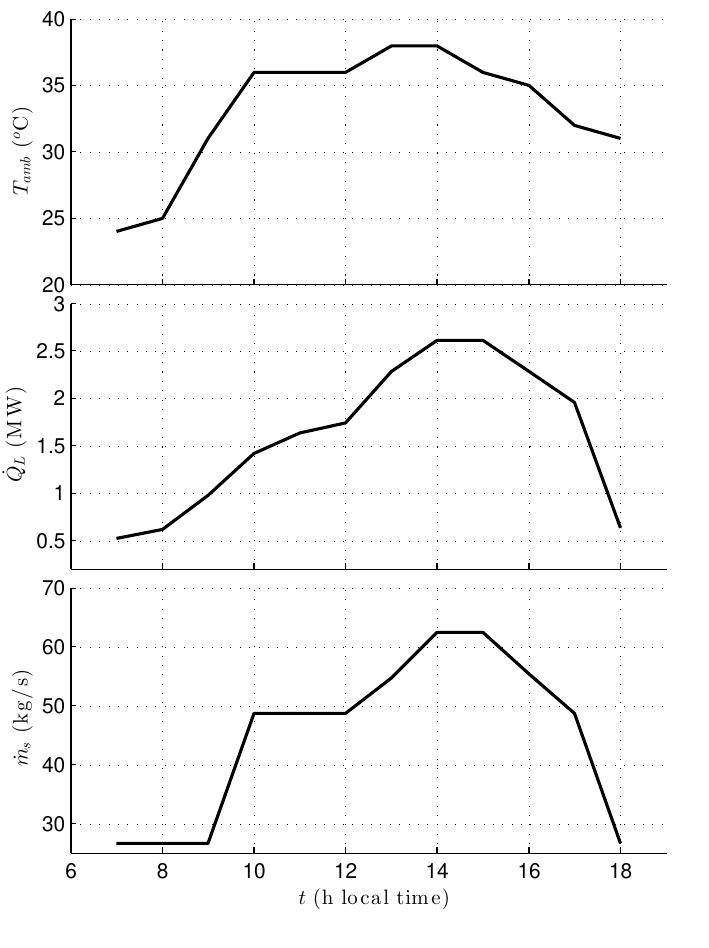}, 
    \includegraphics[width=6.6cm]{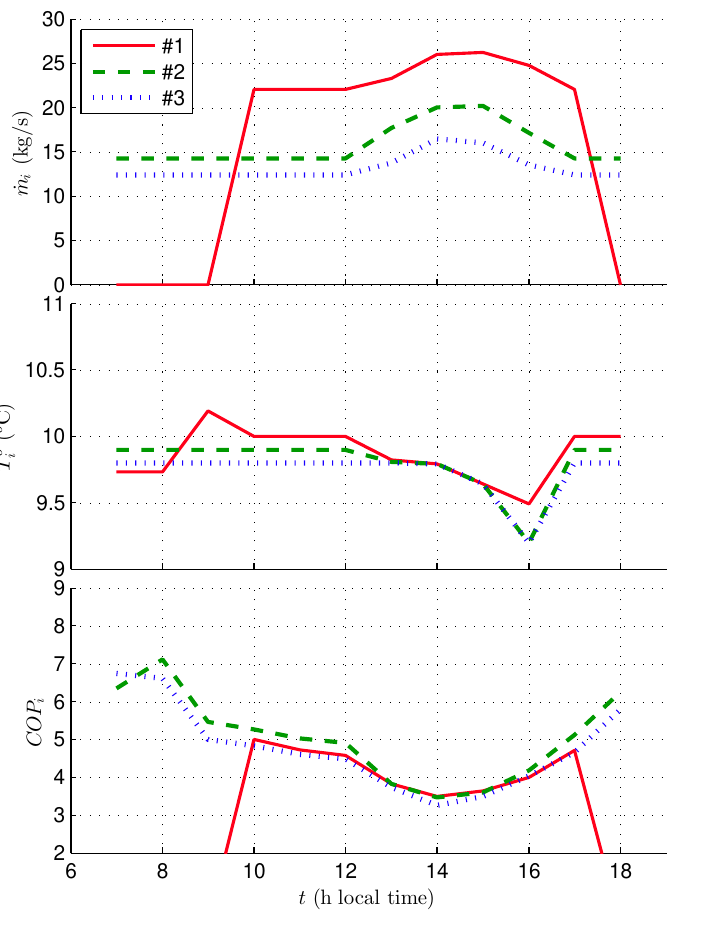}
    \caption{Trajectories for scenario TL1. \textcolor{black}{Legend applies to all graphs.}}
    \label{fig_res_LT1}
\end{figure}

\begin{figure}
    \centering
    \includegraphics[width=6.6cm]{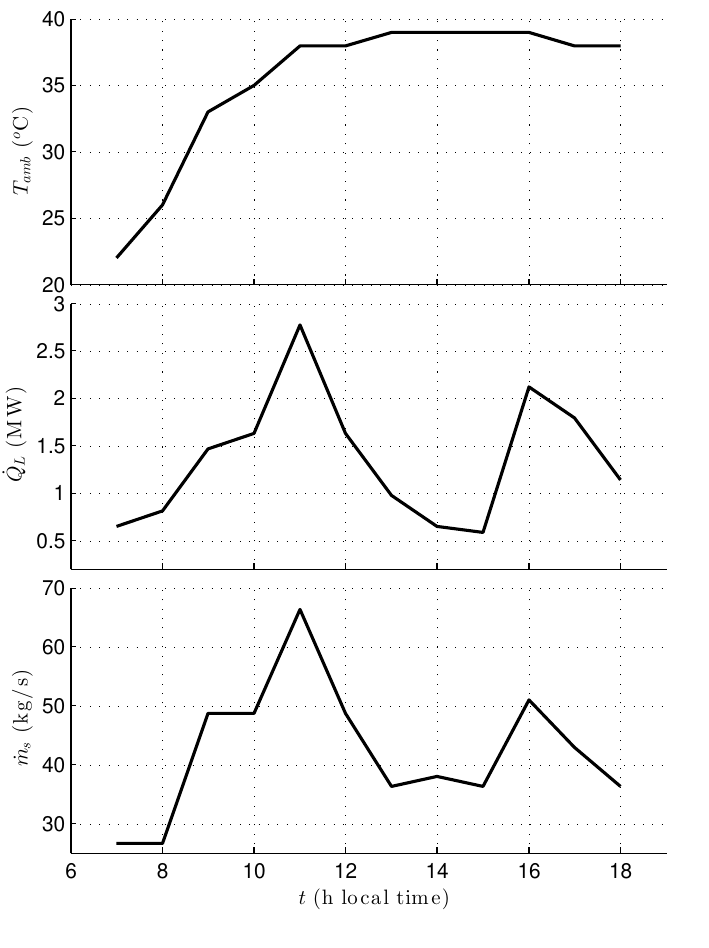}, 
    \includegraphics[width=6.6cm]{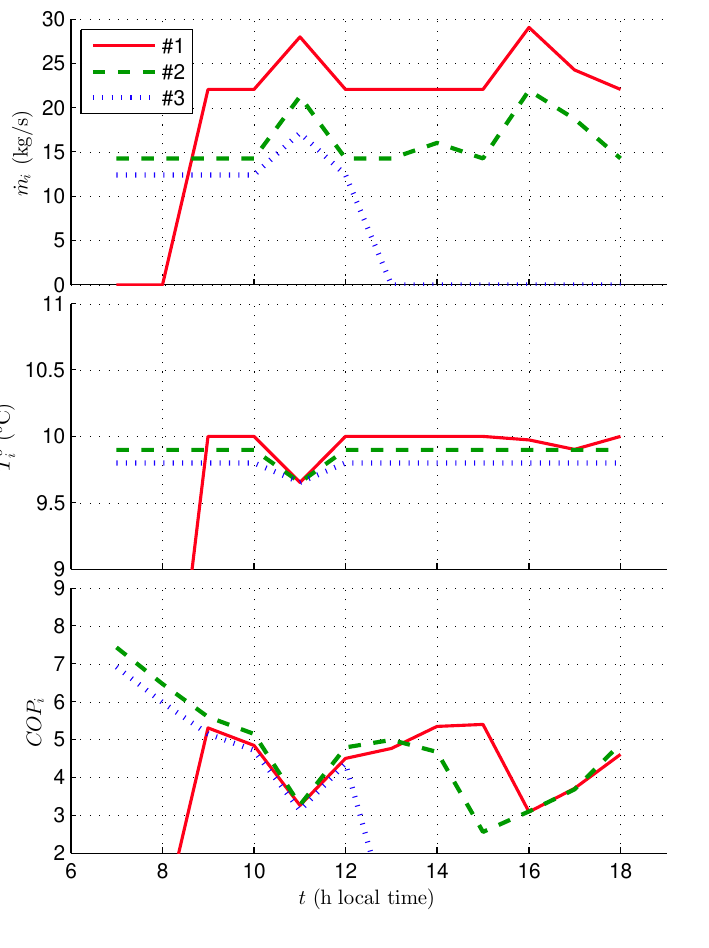}
    \caption{Trajectories for scenario TL2. \textcolor{black}{Legend applies to all graphs.}}
    \label{fig_res_LT2}
\end{figure}

\begin{figure}
    \centering
    \includegraphics[width=6.6cm]{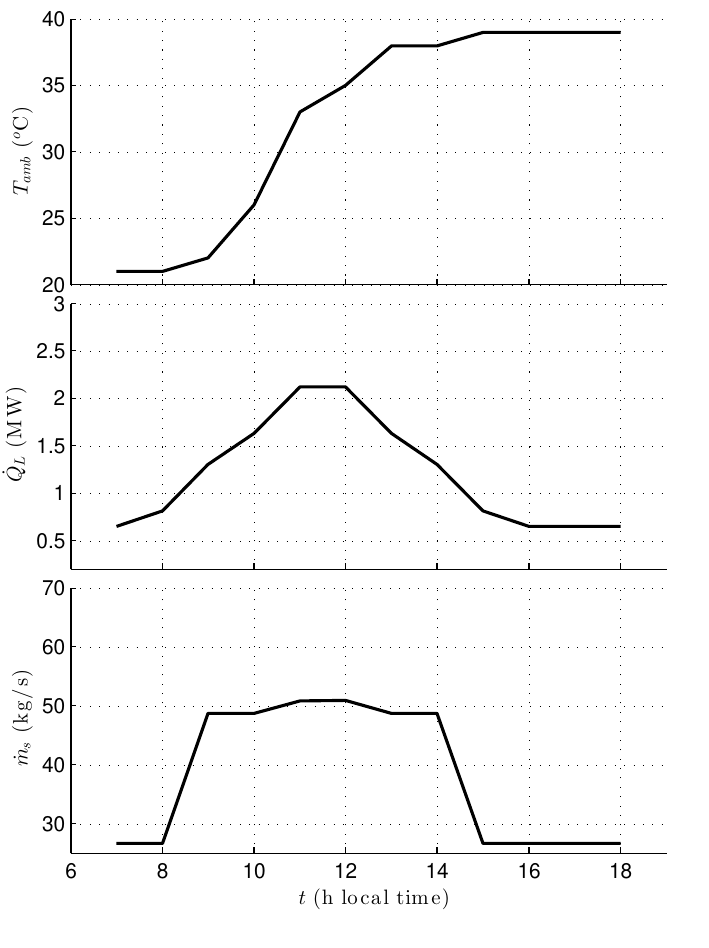}, 
    \includegraphics[width=6.6cm]{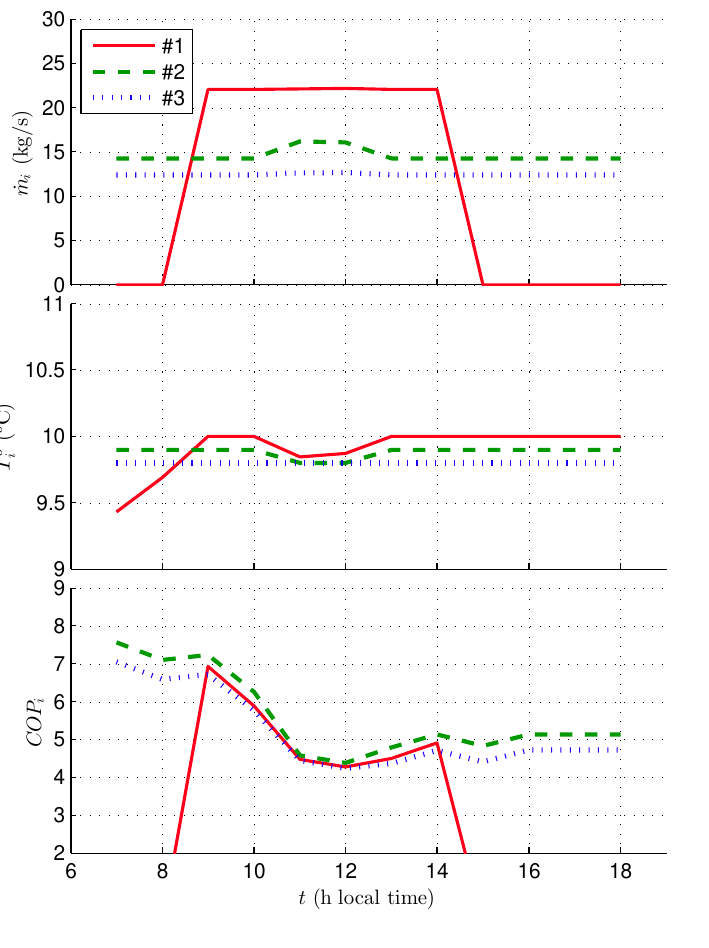}
    \caption{Trajectories for scenario TL3. \textcolor{black}{Legend applies to all graphs.}}
    \label{fig_res_LT3}
\end{figure}

\textcolor{black}{Figures \ref{fig_res_LT1}  to  \ref{fig_res_LT3}  show the solutions for each scenario obtained by the proposed method. The trajectories of decision variables included in ${\bf x}$ are shown in several graphs. The values shown are discrete-time measurements using the hourly resolution (i.e. sampling time $T_s$ is one hour).} The particular ambient temperature trajectory and load profile are also given in separate graphs. \textcolor{black}{In all cases the variables pertaining to chiller 1 (model 400STD) are shown in red solid line, the variables of chiller 2 (model 300STD) in green dashed line, and variables for chiller 3 (model 250STD) in blue dotted line.} Some discussion is provided below.

\begin{enumerate}
    \item TL1. This scenario is characterized by a relatively high temperature and an occupancy with not much fluctuations producing a cooling demand $\dot{Q}_L$ with a peak around 2pm. It can be seen in Fig. \ref{fig_res_LT1} that the optimizer has found most adequate to maintain chillers 2 and 3 working during the 12 hours. Unit 1, however, is allowed to switch on and off. This result is a bit counter-intuitive since the load is a significant one for the day. One would have guessed that maintaining in operation unit 1 (biggest machine in the plant) would be a better option. Table \ref{tab_result} provides a comparison for these two strategies showing that the one adopted by the SPSA algorithm is superior. Also notice that, in addition to turning chiller 1 on and off, this solution uses this same chiller to modulate the output temperature as shown in the graph of $T^o_i$.
    \item TL2. This scenario is characterized by a higher temperature and an occupancy with notable  fluctuations. It can be seen in Fig. \ref{fig_res_LT2} that the cooling demand has two peaks (morning and afternoon). The trajectories of $\dot{m}_i$ show that unit 1 is turn on in the morning and unit 3 is shut down after the first peak, so just two switches take place. Again, from Table \ref{tab_result}  it can be seen that the static solution produces more commutations and a higher consumption.
    \item TL3. This scenario corresponds to a half-holiday and thus occupancy is restricted (for the most part) to the first half of the day. It can be seen in Fig. \ref{fig_res_LT3}  that the solution found consists again in modulating the cooling production with the biggest machine (unit 1) with just two commutations. In Table \ref{tab_result} it is interesting to note that this particular scenario makes the static OCS to produce the highest number of commutations in the state of the chillers.
\end{enumerate}

\textcolor{black}{In order to compare the results of the proposed dynamic OCS with previous approaches, a static OCS has been considered. For clarity the proposed method will be referred to as M1 and the static OCS as M2. The static OCS follows the methods described in \cite{chang2005genetic}; however, the set of decision variables is taken to be the same as in M1 to provide a fair ground for comparison, similarly to \cite{karami2018particle}. These variables provide additional degrees of freedom for improving the performance and, hence, constitute an advantage. Please notice that many papers dealing with OCS consider fixed mass flow rates, thus M2 is a bit more elaborate than those previous approaches. However, M2 will not use dynamic effects in the objective function. As a result M2 can be seen as a classic OCS solving (\ref{eq_optimz_probl}) with $P_e=0$.} 

Table \ref{tab_result} provides a comparison for M1 vs. M2 for the three scenarios introduced above. \textcolor{black}{It can be seen that the number of commutations  for chiller units ($Nc$) is always less for the proposed method (M1). Also, the total consumption $E = \sum_{t=1}^{t=N_T} P \times T_s$ (kWh) for each case is presented. It can be seen that in all cases the  proposed algorithm uses less electrical consumption for the cooling.}

\begin{table}[htp]
    \centering
    \begin{tabular}{p{1.9cm}p{2.3cm}p{2.3cm}p{2.3cm}p{2.4cm}}
    Method & TL1      &   TL2    &  TL3    & AET  \\ \hline \\
    M1     & $Nc = 2$ & $Nc = 2$ & $Nc = 2$ & $AET_1 = 372$ \\
           & $E=5665$ & $E=5136$ & $E=3890$ & $AET_2 = 354$  \\[1mm]
    M2     & $Nc = 3$ & $Nc = 3$ & $Nc = 4$ & $AET_1 = 369$  \\
           & $E=6325$ & $E=5281$ & $E=4271$ & $AET_2 = 359$  \\
    \end{tabular}
    \caption{\textcolor{black}{Results for the proposed dynamic OCS method (M1) and for an enhanced but static OCS method (M2).}}
    \label{tab_result}
\end{table}

\textcolor{black}{The computation time for the algorithm must also be considered in the comparison. The right-most column in Table \ref{tab_result} show the Average Execution Time (AET) in seconds, for the two algorithms using two different hardware types. The first type  correspond to a desktop computer using a series 4 i7 Intel processor with windows 10 operating system and produces the $AET_1$ values. The second type ($AET_2$) is an AMD embedded industrial PC with a AMD Ryzen 3 1200 processor, yielding the $AET_2$ value. It can be seen that the proposed method slightly increases the execution time. Also notice that no real-time requirements are needed due to the long time-scales considered in the problem (sampling time of 1 hour).}

\section{Conclusion}
\textcolor{black}{The research presented in the paper deals with existing gaps in the literature regarding cooling plants with multiple chiller units. In particular, transients have been show to produce additional costs that have not been previously considered in this way. Instead of a fixed penalization for the switching operations, the dynamic effects associated with the transients are modelled. In addition, the mass flow rate and the output temperature reference for each chiller are used as independent variables unlike most OCL/OCS formulation. The resulting problem has been shown to be a nonlinear a constrained optimization.} A Simultaneous Perturbation Stochastic Approximation solution to the dynamic OCS problem has been deployed. The results show that the consideration of transients in the objective function makes the solution of the optimization problem to contain fewer commutations in state of chillers, reducing the transient and associated costs. \textcolor{black}{These results are aligned with the research efforts aiming at reducing energy consumption for cooling. The impact can be significant, specially for large installations.}

\textcolor{black}{The authors believe that future research directions can stem from the introduction of new optimization algorithms that might be better fitted for this kind of problem. Also, other modelling techniques can be applied to derive the relationships needed for the optimization problem formulation.}

\section*{Funding acknowledgement} This research has been funded as Proyecto
RTI2018-101897-B-I00 by FEDER/Ministerio de Ciencia e Innovación –
Agencia Estatal de Investigación. Spain.

\appendix
\section{Nomenclature}
Table \ref{tab_nomenclature} presents a list of the main variables and symbols used throughout the paper.

\begin{table}[tbp]
    \centering
    \begin{tabular}{ll}
    Variables     &  Meaning and units \\ \hline
    $AET$         &  Average Execution Time (s) \\
    $COP$         &  Coefficient of performance \\
    $c_p$         &  Specific heat of water ($W  kg^{-1} K^{-1}$) \\
    $E$           &  Total energy consumption ($kWh$) \\
    $J$           &  Jacobian of objective function \\
    $\dot{m}$     &  Mass flow rate ($kg  s^{-1}$) \\
    $\dot{Q}$     &  Thermal power ($W$) \\
    $P$           &  Electrical power ($W$) \\
    $PLR$         &  Part load ratio ($pu$) \\
    $T$           &  Temperature ($K$) \\
    $T_s$         &  Sampling period (hours) \\
    ${\bf x}$     &  Vector of independent variables \\ \\
    Subscripts    &  Meaning \\ \hline
    $amb$         &  Ambient \\
    $b$           &  Bypass \\
    $i$           &  Chiller number \\
    $j$           &  Iteration \\
    $k$           &  Index to elements in ${\bf x}$ \\
    $L$           &  Load \\
    $l$           &  Thermal losses \\
    $o$           &  Occupancy \\
    $p$           &  Primary \\
    $s$           &  Secondary \\ \\
    Superscripts  &  Meaning \\ \hline
    $co$          &  Condenser \\
    $ev$          &  Evaporator \\
    $i$           &  Input \\
    $o$           &  Output \\
    $n$           &  Nominal \\ \\
    Quantities    &  Meaning \\ \hline
    $N$           &  Number of chiller units \\
    $Nc$          &  Number of commutations \\
    $N_T$         &  Number of time instants \\
    \end{tabular}
    \caption{List of variables and symbols.}
    \label{tab_nomenclature}
\end{table}

 \bibliographystyle{elsarticle-num} 
 \bibliography{cas-refs}





\end{document}